\documentclass[prc,twocolumn,floatfix,groupedaddress,nofootinbib,preprintnumbers,amsmath,amssymb,amsfonts,superscriptaddress,widetable,href]{revtex4}
\usepackage{amsmath,amssymb,amsfonts}
\usepackage{hyperref}
\usepackage{ulem}
\usepackage{graphicx}
\usepackage{color}
\usepackage{longtable}

\allowdisplaybreaks
\begin{document}

\title{Correlations among symmetry energy elements in Skyrme models}
\author{C. Mondal}
\email{chiranjib.mondal@icc.ub.edu}
\address{Saha Institute of Nuclear Physics, 1/AF Bidhannagar, Kolkata
{\sl 700064}, India}
\address{Departament de F\'isica Qu\`antica i Astrof\'isica and Institut de
Ci\`encies del Cosmos (ICCUB), Facultat de F\'isica, Universitat de Barcelona, Mart\'i
i Franqu\`es 1, E-08028 Barcelona, Spain}
\author{B. K. Agrawal}
\email{bijay.agrawal@saha.ac.in}
\address{Saha Institute of Nuclear Physics, 1/AF Bidhannagar, Kolkata
{\sl 700064}, India}
\address{Homi Bhabha National Institute, Anushakti Nagar, Mumbai 400094, India.}
\author{J. N. De}
\email{jn.de@saha.ac.in}
\address{Saha Institute of Nuclear Physics, 1/AF Bidhannagar, Kolkata
{\sl 700064}, India}
\author{S. K. Samaddar}
\email{santosh.samaddar@saha.ac.in}
\address{Saha Institute of Nuclear Physics, 1/AF Bidhannagar, Kolkata
{\sl 700064}, India}

\begin{abstract}

 Motivated by the interrelationships found  between the various symmetry
energy elements of the energy density functionals (EDF) based on  the
Skyrme forces, possible correlations among them are explored.  A total of
237 Skyrme EDFs are used for this purpose. As some of these EDFs yield
values of a few nuclear observables far off from the present acceptable
range, studies are done also with a subset of 162 EDFs that comply
with a conservative set of constraints on the values of  nuclear matter
incompressibility coefficient,  effective mass of the nucleon and the
isovector splitting of effective nucleon masses to see the enhancement
of the correlation strength, if any. The curvature parameter $K_{sym}^0$
and the skewness parameter $Q_{sym}^0$ of the symmetry energy are found
to be very well correlated with the linear combination of the symmetry
energy coefficient and its density derivative $L_0$.  The isovector
splitting of the effective nucleon mass, however, displays a somewhat
meaningful correlation with a linear combination of the symmetry energy,
its slope and its curvature parameter.

\end{abstract}

\maketitle
\section{Introduction}

Lack of accurate knowledge of  the density dependence of nuclear
symmetry energy hinders the understanding of the neutron-rich nuclei
near drip-line \cite{Myers80}.  This knowledge is also of seminal
importance in astrophysical context. The interplay of gravitation with
the pressure of neutron matter (related to the density derivative of
symmetry energy) is a key determining factor in the radii of neutron
stars \cite{Chen14}. The dynamical evolution of the core-collapse
of a massive star and the associated explosive nucleosynthesis
depend sensitively on the density content of the symmetry energy
\cite{Steiner05,Janka07,Oertel17}.  Density dependence of symmetry
energy controls the nature and stability of different phases within a
neutron star, its critical composition, thickness, frequencies of crustal
vibration \cite{Steiner08,Oertel17} and also determines the feasibility
of direct Urca cooling processes within its interior \cite{Todd-Rutel05,
Lattimer91, Steiner05}. The density dependence of symmetry energy
$C_2(\rho)$ around the saturation density can be well expressed in terms
of its slope $L_0$, curvature $K_{sym}^0$ and skewness $Q_{sym}^0$
evaluated at the saturation density ($\rho_0\simeq0.16$ fm$^{-3}$).
These quantities have received a great deal of attention in recent times
\cite{Centelles09,Agrawal12,Moller12,Steiner12,Mondal15,Mondal16,Raduta18}.
Another quantity of topical interest is the isovector splitting of
effective nucleon masses, $\Delta m_0^*$ ( measure of the difference
between neutron and proton effective masses) for asymmetric nuclear
matter defined at $\rho_0$. The value of $\Delta m_0^*$ is very uncertain
\cite{Li13,Li15b,Zhang16,Kong17}.
There is even not much clarity about its sign. Whereas microscopic ab-initio
models consistently predict $\Delta m_0^* > 0$ in neutron-rich matter
\cite{Hassaneen04, Ma04, Dalen05}, parameters of recently
suggested 'best-fit' Skyrme energy density functionals (EDF) \cite{Brown13}
obtained from constraints provided by properties of nuclear matter, of
doubly magic nuclei and microscopic calculations of low-density
neutron matter are found to yield negative values for the isospin-splitted
nucleon effective mass.

The value of $C_2^0$ ($\equiv C_2(\rho_0)$) is known to lie in the range
$\sim 32\pm2$ MeV \cite{Moller12, Myers66, Moller95, Pomorski03, Xu10,
Jiang12}. Extensive efforts have also been made in the last decade
or so to constrain the value of $L_0$ \cite{Centelles09, Agrawal12,
Moller12, Mondal15, Mondal16, Lattimer16, Birkhan17}. The uncertainties
in the values increase further for higher order density derivatives of
symmetry energy i.e. $K_{sym}^0$ or $Q_{sym}^0$. The value of $K_{sym}^0$
ranges from $\sim -700$ MeV to 400 MeV and $Q_{sym}^0$ from $\sim -800$
MeV to 1500 MeV \cite{Dutra12, Dutra14} across a few hundred models of
mean-field energy density functionals (EDF).  There is also enormous
diversity in the predicted values of $\Delta m^*_0$ \cite{Zuo05, Dalen05,
Ou11, Sellahewa14, Chen09, Kong17, Agrawal17, Mondal17}. Comprehensive
understanding of the isovector part of nuclear interaction is thus
hindered. The uncertainties in these nuclear matter constants can,
however, be reduced if one can express them in terms of quantities that
are known in better constraints. Searching for correlated structures
among different symmetry energy elements like $C_2^0, L_0, K_{sym}^0,
Q_{sym}^0$ and $\Delta m^*_0$ is thus highly desirable. The values of
$C_2^0$ and $L_0$ being relatively better established, in recent years,
attempts are made to look for the correlation between $K_{sym}^0$ and
$L_0$ \cite{Danielewicz09, Vidana09, Ducoin11, Tews17}. Apparently,
the correlation shows some degree of model dependence. In a nearly
model independent framework it was, however, analytically shown that
$K_{sym}^0$ is very neatly tied to ($3C_2^0 - L_0$) \cite{Mondal17}. A
strong correlation among them was found using a total of 500 EDFs
based on Skyrme functionals, EDFs based on realistic interactions and
relativistic mean field (RMF) models.  The correlation of $Q_{sym}^0$
with $(3C_2^0-L_0)$ was, however, found to be poor.

The so-found correlation or its absence calls for the need to bring into 
the focus  the analytical relationship among the different symmetry
elements in the EDFs so used. For the EDFs based on realistic interactions
or those based on RMF, finding analytical relationship between different
symmetry elements is not easy, the structure of the Skyrme EDF, however,
 makes it more amenable towards that aim. We try to find that out
in this paper. Once that is done, we explore the solidity of the
correlation between the symmetry energy coefficient and its
higher order derivatives and examine in what context the correlation
is better established. Towards that purpose, initially a total of
237 Skyrme EDFs compiled by Dutra {\it et. al. } \cite{Dutra12}
are used, this is later followed by a restricted set selected out of
them that has compliance with some conservative constraints on the 
Skyrme EDFs to see how the nature of the manifested correlation is affected.

The paper is organized as follows. We present the analytical relations for
various symmetry energy parameters obtained within the Skyrme formalism in
Sec. \ref{formalism}. The results for correlations among various symmetry
energy elements are discussed in Sec.  \ref{results}. Conclusions are
drawn in Sec. \ref{conclusion}.

\section{Formalism}\label{formalism}

The energy per particle $e(\rho,\delta)$ of asymmetric nuclear matter (ANM) 
with density $\rho$ and isospin asymmetry $\delta[=(\rho_n-\rho_p)/\rho]$ is given by,
\begin{eqnarray}
\label{e_anm}
e(\rho,\delta)\simeq e(\rho,\delta=0)+C_2(\rho)\delta^2,
\end{eqnarray}
where $e(\rho,\delta=0)$ is the energy per particle for symmetric nuclear 
matter (SNM) and $C_2(\rho)$ is the symmetry energy defined as,
\begin{eqnarray}
\label{C_2}
C_2(\rho)=\frac{1}{2}\left[\frac{\partial^2 e(\rho,\delta)}{\partial\delta^2}\right]_{\delta=0}.
\end{eqnarray}
Energy per particle for SNM has a minimum at the saturation
density $\rho_0$ around which it can be expanded as,
\begin{eqnarray}
\label{e_snm}
e(\rho,0)\simeq e_0+\frac{1}{2}K_0\epsilon^2+\frac{1}{6}Q_0\epsilon^3,
\end{eqnarray}
where $\epsilon=\frac{\rho-\rho_0}{3\rho_0}$ and $e_0$ the energy per particle
of SNM at $\rho_0$. The incompressibility parameter $K_0$ and stiffness parameter $Q_0$
are defined at $\rho_0$ as,
\begin{eqnarray}
\label{prop_snm}
K_0&=&\left.9\rho^2\frac{\partial^2 e(\rho,0)}{\partial\rho^2}\right|_{\rho_0},\nonumber\\
Q_0&=&\left.27\rho^3\frac{\partial^3 e(\rho,0)}{\partial\rho^3}\right|_{\rho_0}.
\end{eqnarray}
Similarly, the symmetry energy coefficient $C_2(\rho)$ can be expanded around the
saturation density $\rho_0$ in terms of different symmetry energy
elements as,
\begin{eqnarray}
\label{C_2_expand}
C_2(\rho)\simeq C_2^0+L_0\epsilon+\frac{1}{2}K_{sym}^0\epsilon^2+\frac{1}{6}Q_{sym}^0\epsilon^3,
\end{eqnarray}
where the symmetry energy parameters $L_0$, $K_{sym}^0$ and
$Q_{sym}^0$ are related to different density derivatives of $C_2(\rho)$
as,
\begin{eqnarray}
\label{prop_sym}
L_0&=&\left.3\rho\frac{\partial C_2(\rho)}{\partial \rho}\right|_{\rho_0},\nonumber\\
K_{sym}^0&=&\left.9\rho^2\frac{\partial^2 C_2(\rho)}{\partial \rho^2}\right|_{\rho_0},\nonumber\\
Q_{sym}^0&=&\left.27\rho^3\frac{\partial^3 C_2(\rho)}{\partial \rho^3}\right|_{\rho_0}.
\end{eqnarray}

In the standard Skyrme parametrization one can write the expression for
energy per particle of asymmetric nuclear matter of density $\rho$
and asymmetry
$\delta$ as \cite{Chabanat97},
\begin{eqnarray}
\label{Chabanat_erho}
e(\rho,\delta)&=&\frac{3}{5}\frac{\hbar^2}{2m}\left(\frac{3\pi^2}{2}\right)^{2/3}
\rho^{2/3}F_{5/3}\nonumber\\
&+&\frac{1}{8}t_0 \rho[2(x_0+2)-(2x_0+1)F_2]\nonumber\\
&+&\frac{1}{48}t_3\rho^{\alpha+1}[2(x_3+2)-(2x_3+1)F_2]\nonumber\\
&+&\frac{3}{40}\left(\frac{3\pi^2}{2}\right)^{2/3}\rho^{5/3}\Big\{[t_1(x_1+2)
+t_2(x_2+2)]F_{5/3}\nonumber\\
&+&\frac{1}{2}[t_2(2x_2+1)-t_1(2x_1+1)]F_{8/3}\Big\},
\end{eqnarray}
where, $F_l(\delta)=\frac{1}{2}\left[(1+\delta)^l+(1-\delta)^l\right]$.
All the parameters $t_i$'s, $x_i$'s etc. can be expressed in terms of
nuclear matter properties. Doing that one observes that the parameters
$t_0, t_3, \alpha$ are completely determined by the bulk properties of
SNM. On the other hand, the other parameters $x_0, x_1, x_2, x_3, t_1,
t_2$ are connected to isovector elements of asymmetric nuclear matter
\cite{Gomez92}.

Following the expression for $C_2(\rho)$ in
Eq. (\ref{C_2}) and the definitions of different symmetry energy elements
in Eq. (\ref{prop_sym}) one can write the expressions for them in the Skyrme
formalism as (see Appendix \ref{appendix1}),
\begin{eqnarray}
\label{skyrme_sym}
C_2^0&=&\frac{1}{3}E_F^0
-\frac{1}{8}t_0(2x_0+1)\rho_0\nonumber\\
&-&\frac{1}{24}\left(\frac{3\pi^2}{2}\right)^{2/3}\Big(3t_1x_1-t_2(4+5x_2)
\Big)\rho_0^{5/3}\nonumber\\
&-&\frac{1}{48}t_3(2x_3+1)\rho_0^{\alpha+1},\\
L_0&=&\frac{2}{3}E_F^0
-\frac{3}{8}t_0(2x_0+1)\rho_0\nonumber\\
&-&\frac{5}{24}\left(\frac{3\pi^2}{2}\right)^{2/3}\Big(3t_1x_1-t_2(4+5x_2)
\Big)\rho_0^{5/3}\nonumber\\
&-&\frac{1}{16}(\alpha+1)t_3(2x_3+1)\rho_0^{\alpha+1},
\end{eqnarray}
\begin{eqnarray}
\label{skyrme_ksym}
K_{sym}^0&=&-5(3C_2^0-L_0)+E_F^0\nonumber\\
&+&\frac{1}{16}\alpha(2-3\alpha)t_3(2x_3+1)\rho_0^{\alpha+1},\\
\label{skyrme_qsym}
Q_{sym}^0&=&(3\alpha+2)K_{sym}^0+15(\alpha+1)(3C_2^0-L_0)\nonumber\\
&-&(3\alpha+1)E_F^0.
\end{eqnarray}
Here, $E_F^0$ is the Fermi energy of SNM at $\rho_0$ given by,
$E_F^0=\frac{\hbar^2}{2m}\left(\frac{3\pi^2}{2}\right)^{2/3}\rho_0^{2/3}$.
The effective mass $m_q^*$ of a nucleon [$q=n/p$ (neutron/proton)], at 
density $\rho_0$ in the Skyrme formalism is given from the relation 
\cite{Chabanat98}
\begin{eqnarray}
\label{eff1}
\frac{\hbar^2}{2m_q^*}&=&\frac{\hbar^2}{2m}+\frac{1}{8}[t_1(2+x_1)+t_2(2+x_2)]\rho_0
\nonumber\\
&-&\frac{1}{8}[t_1(1+2x_1)-t_2(1+2x_2)]\rho_q.
\end{eqnarray}
The isospin-splitted effective nucleon 
mass is defined as the difference between the neutron and proton effective masses, 
at $\rho_0$ it is 
\begin{eqnarray}
\label{delm_def}
\Delta m^*_0=\left.\left[\frac{m_n^*-m_p^*}{m}\right]_{\rho_0}\right/\delta.
\end{eqnarray}
In Skyrme formalism it is written in terms of the symmetry elements 
$C_2^0$, $L_0$ etc. as
\begin{eqnarray}
\label{delm2}
&&\Delta m_0^*=\Bigg(K_{sym}^0+3(1+\alpha)(3C_2^0-L_0)+(1-3\alpha)E_F^0\nonumber\\
&&+\frac{2}{3}(3\alpha-2)\frac{m}{m_0^*}E_F^0\Bigg)\Bigg/\left[(3\alpha-2)E_F^0\left(
\frac{m}{m_0^*}\right)^2\right].
\end{eqnarray}
Some details for the derivation of Eqs. (\ref{skyrme_ksym}-\ref{delm2}) are 
given in the Appendix \ref{appendix1}.

\section{Results and Discussions}\label{results}

Using relations among different thermodynamic state functions, without
any specific form of the nuclear interaction but with only viable
approximations on its nature, in a recent work \cite{Mondal17} it was
shown that a nearly universal correlation exists between $K_{sym}^0$ and
$(3C_2^0-L_0)$. Imposing a general constraint that the neutron energy per
particle should be zero at zero density of neutron matter, a plausible 
explanation of such a correlation was recently given \cite{Margueron18}. 
The structure of the Eqs. (\ref{skyrme_ksym}-\ref{delm2})
suggests that such a correlated structure among the different symmetry
energy elements might also exist in the Skyrme EDF framework. With this
in mind, we have performed an analysis using all the Skyrme EDFs compiled
by Dutra {\it et. al.} \cite{Dutra12} except the ones namely ZR3a, ZR3b
and ZR3c \cite{Jaqaman84}, where the symmetry energy $C_2^0$ is negative.
We present the results obtained using all the 237 Skyrme EDFs, referred to
as `ALL' hereafter.  Some of these EDFs, however, yield values of nuclear
constants like the nuclear incompressibility $K_0$ and the nucleon
effective mass $m_0^*$ of SNM beyond present acceptable range. Estimates
of $K_0$ obtained from analyzes related to isoscalar giant monopole
resonances (ISGMR) \cite{Colo92, Blaizot95, Hamamoto97, Niksic08, Khan12,
Avogadro13, De15, Shlomo06} is now well constrained to $K_0 =230\pm 30
$ MeV.  The effective mass $m_0^*$ varies between $\sim0.7m$ \cite{Li15b,
Jaminon89} to $\sim1.1m$ \cite{Kortelainen10,Kortelainen12,Kortelainen14}.
Experimental and theoretical studies of isoscalar giant quadrupole
resonance (ISGQR) \cite{Bohigas79, Klupfel09, Roca-Maza13, Zhang16}
suggest a value of $\frac{m_0^*}{m}\simeq 0.90$, but more experimental
data may be needed for a better quantification.  We choose to constrain
it at $\frac{m_0^*}{m} =0.85 \pm 0.15 $.  Imposing a  further constraint
on the isovector splitting of effective mass $|\Delta m_0^*| < 1$
(which more than covers the values from the limited experimental data
\cite{Zhang16,Coupland16,Kong17} and recent theoretical values on it),
the restricted set of Skyrme EDFs is downsized to 162 in number. This
is referred to as the `SELECTED' set. Calculations are performed with
this selected set also to see how the correlations are affected.

\begin{figure}[h]{}
\includegraphics[height=3.4in,width=3.2in,angle=-90]{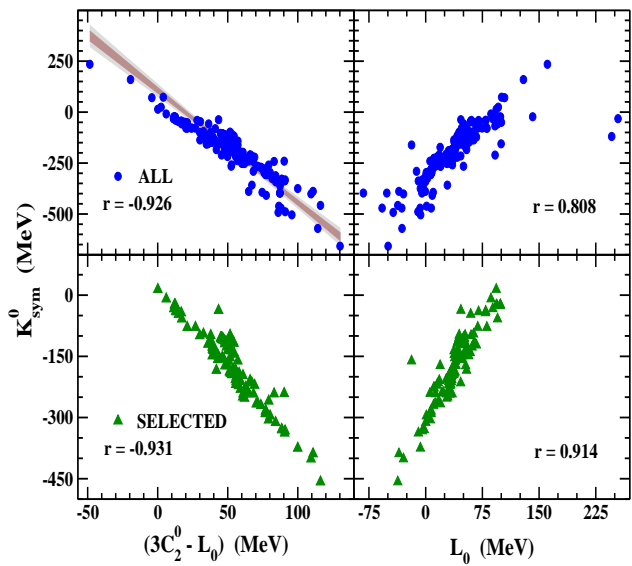}
\caption{\label{fig1}
(Color online)  The correlation of $K_{sym}^0$ with $L_0$ and with
$[3C_2^0-L_0]$ are depicted in the right and left panels, respectively.
Results for 237 Skyrme EDFs (`ALL') are displayed in the upper panels,
the lower panels contain the results for a selected subset of 162 models
(`SELECTED') (see text for details).  The inner(outer) colored regions
around the best-fit straight line in upper left panel depict the loci
of 95 $\% $ confidence (prediction) bands of the regression analysis. }
\end{figure}

Correlations between second and higher order density derivatives of
symmetry energy with the slope parameter $L_0$ have earlier been studied
in the literature \cite{Danielewicz09,Ducoin11,Tews17} with some Skyrme
EDFs. The results are mixed, the degree of correlation is found to depend
on the subjective choice of selection of models.  Eq. (\ref{skyrme_ksym})
however shows  that $K_{sym}^0$ may be better correlated with
$(3C_2^0-L_0)$. The correlation plot between them for all the 237 Skyrme
EDFs is displayed in the upper left panel of Fig. \ref{fig1}. A linear
correlation as suggested in Eq. (\ref{skyrme_ksym}) is observed, the
correlation coefficient is $r=-0.926$. As mentioned earlier, $K_{sym}^0$
is a poorly determined quantity, existence of this correlation enables one
to determine its value from relatively better known isovector quantities
$C_2^0$ and $L_0$ by using the best fit straight line $K_{sym}^0 =
a(3C_2^0-L_0)+c$ with $a=-5.51\pm0.15$ and $c=106.84\pm3.37$ MeV. One
notes that $a$ is not too far from -5, the coefficient of $(3C_2^0-L_0)$
in Eq. (\ref{skyrme_ksym}).  In Ref. \cite{Ducoin11}, for a restricted
set of EDFs, the correlation of $K_{sym}^0$ with $L_0$ was studied and
reported to be strong.  For the set 'ALL' of Skyrme EDFs, we find it to be
weaker ($r$=$0.808$). This correlation is shown in the upper right panel
of Fig.\ref{fig1}.  The correlation between $K_{sym}^0$ and $(3C_2^0-L_0)$
is then tested for the `SELECTED' set. It increases marginally, however,
a marked improvement in the correlation between $K_{sym}^0$ and $L_0$
is observed.  These correlations are displayed in the bottom panels of
Fig. \ref{fig1}.  The correlation of $K_{sym}^0$ with $(3C_2^0-L_0)$
is thus seen to be more robust compared to that with $L_0$.

\begin{figure}[h]{}
\includegraphics[height=3.4in,width=3.2in,angle=-90]{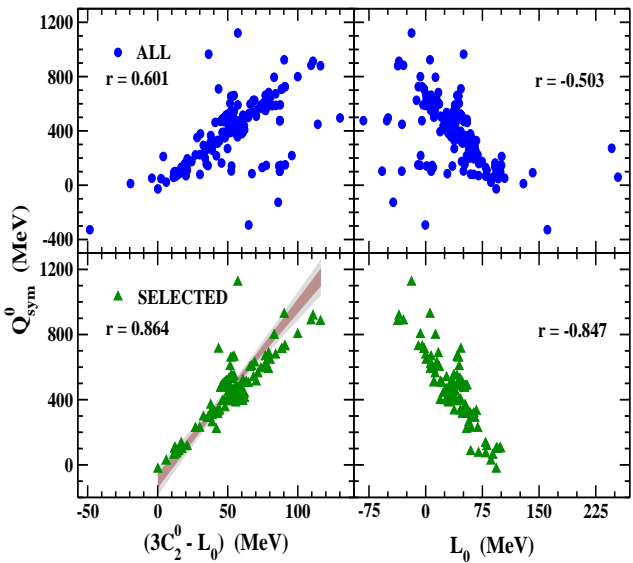}
\caption{\label{fig2}
(Color online)  Same as Fig \ref{fig1} but for $Q_{sym}^0$. The confidence bands
of regression analysis are given only for the subset of models (`SELECTED') 
in the lower left panel.}
\end{figure}

The plots of $Q_{sym}^0$ as a function of $(3C_2^0-L_0)$
and of $L_0$ for the set 'ALL' is shown in the upper panels of
Fig. \ref{fig2}. $Q_{sym}^0$ is seen to be poorly correlated
with $L_0$. The situation does not improve
significantly for the correlation of $Q_{sym}^0$ with $(3C_2^0 - L_0)$.
Eq. \ref{skyrme_qsym} shows $Q_{sym}^0$ to be nearly a linear combination
of $K_{sym}^0$ and $(3C_2^0-L_0)$ and $K_{sym}^0$ is seen to be well
correlated with $(3C_2^0-L_0)$; one may thus expect $Q_{sym}^0$ to
be well correlated with $(3C_2^0-L_0)$.  A weak correlation is found
though between $Q_{sym}^0$ and $(3C_2^0-L_0)$ with all the models
with a correlation coefficient $r=0.6$.  However, with the subset
of models (`SELECTED'), substantial improvement in the correlation
between $Q_{sym}^0$ and $(3C_2^0-L_0)$ can be observed ($r=0.864$).
Due to imposed constraints on $K_0$ and $\frac{m_0^*}{m}$, the values of
$\alpha$ for the Skyrme models (as used in the formalism) get limited
to a narrower range. That is why the correlation between $Q_{sym}^0$
and $(3C_2^0-L_0)$ improves significantly for the  `SELECTED' set.
Even the correlation of $Q_{sym}^0$ with $L_0$ shows a marked gain
in this case as displayed in the right bottom panel; it is nearly the
same as with $(3C_2^0-L_0)$. Correlation of $Q_{sym}^0$ with a linear
combination of $K_{sym}^0$ and $(3C_2^0-L_0)$ is seen to be quite robust
for both the sets, with correlation coefficient $r \sim 0.95$ (not shown
in the figure). This robustness points to the fact that even though
a strong correlation exists between $K_{sym}^0$ and $(3C_2^0-L_0)$,
a minute deviation from exact correlation may affect the correlation
with higher order derivatives.

\begin{table*}[t]
\caption{\label{tab1}The fitted expressions for $K_{sym}^0$, $Q_{sym}^0$
and $\Delta m_0^*$ along with the fitted parameters are listed. In
the third column `A' means `ALL' the 237 EDFs which are employed for
the analysis and `S' means the `SELECTED' set  that are chosen where
$K_0=230\pm30$ MeV, $\frac{m_0^*}{m}=0.85\pm0.15$ and $|\Delta m_0^*|<1$.
The units of the coefficients  $a, b $ and $c$ are such that they yield
the values of $K_{sym}^0$ and $Q_{sym}^0$ in MeV and $\Delta m_0^*$ comes
out to be dimensionless.  The correlation coefficient `$r$' is listed in
the seventh column. The last column shows the estimates of $K_{sym}^0$,
$Q_{sym}^0$ and $\Delta m_0^*$ along with their uncertainties once $C_2^0$
and $L_0$ are given. The numbers in the parentheses depict the estimates
once the dispersions in $C_2^0$ and $L_0$ are included.}

 \begin{ruledtabular}
\begin{tabular}{cccccccc}
Quantity  & fitted expression & Set  & $a$ & $b$ & $c$ &
$r$ & estimate  \\
\hline
$K_{sym}^0$ & $a (3C_2^0-L_0)+c$ & A & $-5.51\pm0.15$ && $106.84\pm3.37$ 
 & -0.926 & $-97.0\pm6.5$(86.8)\\
& $a (3C_2^0-L_0)+c$ & S& {$-4.56\pm0.14$} && {$71.80\pm2.70$} 
 & {-0.931} & $-96.9\pm5.8$(71.9)\\
$Q_{sym}^0$ & $a (3C_2^0-L_0)+c$ & S& {$10.72\pm0.49$} && {$-97.38\pm8.84$} 
 & {0.864} & $299.4\pm20.2$(169.2)\\
& $a K_{sym}^0+b (3C_2^0-L_0)+c$ & A & $3.51\pm0.07$ & $22.21\pm0.42$ &
$-115.87\pm6.03$ & 0.952 & $365.4\pm28.0$(147.3)\\
& $a K_{sym}^0+b (3C_2^0-L_0)+c$ & S& {$2.65\pm0.05$} & {$18.88\pm0.38$} &
{$-86.87\pm3.73$} & {0.965} & $354.5\pm22.3$(139.1)\\
$\Delta m_0^*$ & $a K_{sym}^0+b (3C_2^0-L_0)+c$ & S& {$-0.0094\pm0.0003$} & {$-0.0363\pm0.0012$} &
{$0.3958\pm0.0202$} & {0.790} & $-0.034\pm0.081$(0.260)\\
\end{tabular}
\end{ruledtabular}
\end{table*}

\begin{figure}[h]{}
\includegraphics[height=3.4in,width=3.2in,angle=-90]{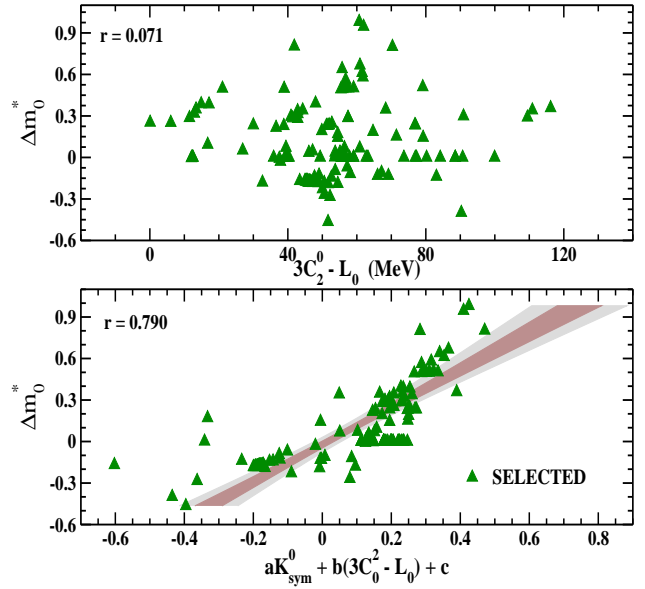}
\caption{\label{fig3}
(Color online) The correlation of $\Delta m_0^*$ with $(3C_2^0-L_0)$ is depicted in the 
upper panel and with $a K_{sym}^0+b(3C_2^0-L_0)+c$ in the lower panel 
with selected set of models (`SELECTED'). The values of $a,\ b$ and $c$ are given 
in the last row of Table \ref{tab1}. The $95\%$ confidence (prediction) band 
is also depicted by brown (grey) region in the lower panel.}
\end{figure}
The symmetry element $\Delta m_0^*$ shows practically no correlation with
either $L_0$ or with $(3C_2^0-L_0)$ for the complete set of EDFs `ALL'.
The situation does not improve with the `SELECTED'
set  which can be noted from the upper panel of Fig. \ref{fig3}.
We therefore looked for a correlation between $\Delta m_0^*$ and a
linear combination of $K_{sym}^0$ and $(3C_2^0-L_0)$ as suggested in
Eq. (\ref{delm2}). Even then, a meaningful correlation could not be
found for the full EDF set ($r=0.286$).  The reason behind this is
that $\Delta m_0^*$ is a sum of small positive and negative numbers
(see Eq. (\ref{delm2})). Small errors in those quantities may shadow
the correlation. That is why,  the selected set with reasonable constraints
on few nuclear matter properties pulls the correlation up to a somewhat
significant value ($r=0.790$). This is shown in the lower panel of Fig.
\ref{fig3}.

Results of the correlation analyses are presented in Table \ref{tab1}. One
notices from the table that the values of the constants appearing in
the best-fit equations connecting one symmetry element with others with
significant correlation are in near harmony with those that appear
in Eq. (\ref{skyrme_ksym}-\ref{delm2}).  For instance,  the slope
'$a$' in the best fit equation $K_{\rm sym}^0 = a(3C_2^0 - L_0) + c$
is found to be $a =-5.51\pm0.15$ for all models and $a =-4.56\pm0.14$
for the selected set of models; this is compatible with the coefficient
of $(3C_2^0 - L_0)$ in Eq. (\ref{skyrme_ksym}). Similar is the case for
the coefficients  '$a$' and '$b$' in the expressions $Q_{\rm sym}^0 =
a K_{\rm sym}^0 +b (3C_2^0 - L_0) +c$ and for $\Delta m_0^*=a K_{\rm
sym}^0 +b (3C_2^0 - L_0)+c$ with $\alpha\sim 0.2$, $E_F^0\sim 36$ MeV and
$\frac{m_0^*}{m}\sim0.85$. From meaningful correlation coefficients as
listed in Table \ref{tab1}, with fiducial values of $C_2^0 = 32$ MeV and
$L_0 = 59$ MeV \cite{Li13}, the values of $K_{sym}^0$ and $Q_{sym}^0$
are seen to be nearly independent of the set chosen.  Their values are
reported in the last column of Table I.
The errors in the calculated quantities arise due to lack of perfect
correlation between the symmetry coefficients. Besides the imperfect
correlation, the uncertainties in the values of $C_2^0 (32\pm 2$ MeV) 
and $L_0 (59 \pm 15$ MeV) cause a considerably large dispersion in 
the values of the symmetry elements. They are also shown in the
parentheses in the last column of Table 1. The value of symmetry
incompressibility $ K_{\tau} (=K_{sym}-6L_0-\frac{Q_0L_0}{K_0})$ can
be estimated provided the value of the skewness parameter $Q_0$ is
known. There is no experimental knowledge on $Q_0$; with the constraint
on the value of $K_0 (230 \pm 30$ MeV), in the selected set of Skyrme
EDFs, $Q_0$ is seen to lie in a very narrow range ($Q_0=-370 \pm 25 $
MeV). The value of $K_\tau $ then turns out to be $K_\tau =-356 \pm 93$
MeV, matching quite well the recent theoretical estimates \cite{Vidana09,
Pearson10, Alam14}.  The isospin-splitted effective mass $\Delta m_0^*$
comes out to be  slightly negative.  This is, however, found to be very
sensitive on the value of $(3C_2^0-L_0)$ indicating that the present
knowledge in the accuracy of $L_0$ may be incapable of extracting $\Delta
m_0^*$ reliably.  Its value may also partly depend on the definition
of effective mass (Eq. \ref{eff1}). In astrophysical context, terms beyond
the linear in density have been suggested \cite{Baldo14} for the
evaluation of the nuclear effective mass, however, at the saturation
density $\rho_0$, their effect on $\Delta m_0^*$ are found to be not
very significant.

\section{Conclusions}\label{conclusion}

Calculations on the correlation between the symmetry coefficients presented
in this paper are done in the ambit of the Skyrme EDFs. Undeniable 
model dependence in the conclusions arrived so far thus can not be 
ruled out. However, knowing that the isovector wing of the nuclear
interaction is not yet very precise, constraining it through a 
structural relationship among the nuclear symmetry elements bearing
its imprint is highly relevant even in a model, particularly when the
model (Skyrme) has been extremely successful in explaining  diverse
experimental data. With this objective, analytical expressions for
the different symmetry energy elements are obtained for the standard
Skyrme EDFs and cast in forms suggestive of correlation between the
lower and higher order density derivatives of symmetry energy. To
this purpose we have employed 237 Skyrme class of energy density
functionals \cite{Dutra12}.  Calculations reveal that there is a robust
correlation between $K_{sym}^0$ and $(3C_2^0-L_0)$.  The calculations
were repeated for a set of restricted EDFs selected with imposition of a
conservative set of empirical constraints on the values of nuclear matter
incompressibility, effective mass of the nucleon and the isospin-splitted
nucleon effective mass.  The insignificant change in the correlation
coefficient reinforces the robustness of the correlation between
$K_{sym}^0$ and $(3C_2^0-L_0)$.  Even if the value of $K_{sym}^0$ varies
widely in the Skyrme EDFs, it can be better bound with provision for good
empirical knowledge of $C_2^0$ and $L_0$. $Q_{sym}^0$ is also reasonably
correlated with $(3C_2^0-L_0)$ but with only the restricted set of EDFs.
A strong correlation of $Q_{sym}^0$ with linear combination of $K_{sym}^0$
and $(3C_2^0-L_0)$ is also observed. For $\Delta m_0^*$, a somewhat
meaningful correlation is found with linear combination of $K_{sym}^0$
and $(3C_2^0-L_0)$ subject to the  restricted set of EDFs. To the best
of our knowledge, strong correlation of $Q_{sym}^0$ and even moderate
correlation of  $\Delta m_0^*$ with other symmetry energy parameters have
not been reported earlier.  The symmetry energy coefficient $C_2^0$ is a
somewhat well estimated quantity extracted from different experimental
observations. Though, $L_0$ is not that well determined as $C_2^0$,
progress in constraining it is going on for some years. Experiments like
PREX-II \cite{Abrahamya12} give promises for a better determination
of $L_0$ in a model independent way. Thus exploiting the correlated
structures we have presented, one can estimate the higher order symmetry
energy derivatives like $K_{sym}^0$ and $Q_{sym}^0$ in good bounds. It
is not that certain for $\Delta m_0^*$. As seen from Eq. (\ref{delm2})
and the last row of the Table \ref{tab1}, it is a sum of small positive
and negative terms; even small errors in the other symmetry elements
may render it very uncertain.

\acknowledgments
J.N.D. is thankful to the Department of Science and Technology, Government of India
for support with the Grant EMR/2016/001512.

\appendix\section{Symmetry energy parameters in Skyrme Formalism}\label{appendix1}
Following the definition (Eq. (\ref{C_2})), the symmetry energy $C_2(\rho)$ 
is obtained from Eq. (\ref{Chabanat_erho}) as 
\cite{Chabanat97},
\begin{eqnarray}
\label{c2}
C_2(\rho)&=&\frac{1}{2}\left[\frac{\partial^2e(\rho,\delta)}{\partial\delta^2}
\right]_{\delta=0}\nonumber\\
&=&\frac{1}{3}\frac{\hbar^2}{2m}\left(\frac{3\pi^2}{2}\right)^{2/3}\rho^{2/3}
-\frac{1}{8}t_0(2x_0+1)\rho\nonumber\\
&-&\frac{1}{24}\left(\frac{3\pi^2}{2}\right)^{2/3}\Big(3t_1x_1-t_2(4+5x_2)
\Big)\rho^{5/3}\nonumber\\
&-&\frac{1}{48}t_3(2x_3+1)\rho^{\alpha+1}.
\end{eqnarray}
Similarly, taking first and second order derivatives of $C_2(\rho)$ with respect 
to $\rho$, expressions for $L(\rho)$ and $K_{sym}(\rho)$ can be obtained 
respectively as,
\begin{eqnarray}
\label{l}
L(\rho)&=&3\rho\left(\frac{\partial C_2(\rho)}{\partial \rho}\right)\nonumber\\
&=&\frac{2}{3}\frac{\hbar^2}{2m}\left(\frac{3\pi^2}{2}\right)^{2/3}\rho^{2/3}
-\frac{3}{8}t_0(2x_0+1)\rho\nonumber\\
&-&\frac{5}{24}\left(\frac{3\pi^2}{2}\right)^{2/3}\Big(3t_1x_1-t_2(4+5x_2)
\Big)\rho^{5/3}\nonumber\\
&-&\frac{1}{16}(\alpha+1)t_3(2x_3+1)\rho^{\alpha+1},\\
\label{ksym1}
K_{sym}(\rho)&=&9\rho^2 \left(\frac{\partial^2 C_2(\rho)}{\partial \rho^2}\right)\nonumber\\
&=&-\frac{2}{3}\frac{\hbar^2}{2m}\left(\frac{3\pi^2}{2}\right)^{2/3}\rho^{2/3}
\nonumber\\
&-&\frac{5}{12}\left(\frac{3\pi^2}{2}\right)^{2/3}\Big(3t_1x_1-t_2(4+5x_2)
\Big)\rho^{5/3}\nonumber\\
&-&\frac{3}{16}\alpha(\alpha+1)t_3(2x_3+1)\rho^{\alpha+1}.
\end{eqnarray}
From Eqs. \ref{c2} and \ref{l} one obtains the expression of $(3C_2^0-L_0)$ as 
\begin{eqnarray}
\label{3c2l}
(3C_2^0-L_0)&=&\frac{1}{3}\frac{\hbar^2}{2m}\left(\frac{3\pi^2}{2}\right)^{2/3}\rho^{2/3}
\nonumber\\
&+&\frac{1}{12}\left(\frac{3\pi^2}{2}\right)^{2/3}\Big(3t_1x_1-t_2(4+5x_2)
\Big)\rho^{5/3}\nonumber\\
&+&\frac{1}{16}\alpha t_3(2x_3+1)\rho^{\alpha+1}.
\end{eqnarray}
Using the expression of $(3C_2^0-L_0)$ in Eq. (\ref{3c2l})  
one obtains the expression for $K_{sym}$ at $\rho_0$ as,
\begin{eqnarray}
\label{ksym2}
K_{sym}^0&=&-5(3C_2^0-L_0)+E_F^0\nonumber\\
&+&\frac{1}{16}\alpha(2-3\alpha)t_3(2x_3+1)\rho_0^{\alpha+1}.
\end{eqnarray}
Here, $E_F^0$ is the Fermi energy of the system at $\rho_0$ given by, 
$E_F^0=\frac{\hbar^2}{2m}\left(\frac{3\pi^2}{2}\right)^{2/3}\rho_0^{2/3}$.

Taking third derivative of $C_2(\rho)$ given in Eq. (\ref{c2}) one can arrive 
at the expression for $Q_{sym}(\rho)$ as,
\begin{eqnarray}
\label{qsym1}
Q_{sym}(\rho)&=&27\rho\left(\frac{\partial^3 C_2(\rho)}{\partial \rho^3}\right)\nonumber\\
&=&\frac{8}{3}\frac{\hbar^2}{2m}\left(\frac{3\pi^2}{2}\right)^{2/3}\rho^{2/3}
\nonumber\\
&+&\frac{5}{12}\left(\frac{3\pi^2}{2}\right)^{2/3}\Big(3t_1x_1-t_2(4+5x_2)
\Big)\rho^{5/3}\nonumber\\
&-&\frac{9}{16}\alpha(\alpha+1)(\alpha-1)t_3(2x_3+1)\rho^{\alpha+1}.
\end{eqnarray}
After simplification one can express $Q_{sym}$ in terms of 
nuclear matter properties at $\rho_0$ as,
\begin{eqnarray}
\label{qsym2}
Q_{sym}^0&=&(3\alpha+2)K_{sym}^0+15(\alpha+1)(3C_2^0-L_0)\nonumber\\
&-&(3\alpha+1)E_F^0.
\end{eqnarray}

To find the expression for $\Delta m_0^*$ we take recourse to Eq. (\ref{eff1}). 
Defining $m_0^*$ as the effective mass for SNM, from Eq. (\ref{eff1}), one obtains
\begin{eqnarray}
\label{eff2}
\frac{\hbar^2}{2m_0^*}-\frac{\hbar^2}{2m}&-&\frac{3\hbar^2}{2}\left(\frac{\hbar^2}{2m_p^*}-\frac{\hbar^2}{2m_n^*}
\right)\nonumber\\
&=&\frac{1}{8}[t_2(4+5x_2)-3t_1x_1]\rho_0.
\end{eqnarray}
Replacing RHS of Eq. (\ref{eff2}) from Eq. (\ref{3c2l}) one can write
\begin{eqnarray}
\label{delm0}
3C_2^0-L_0&=&E_F^0-\frac{2}{3}\frac{m}{m_0^*}E_F^0+E_F^0 m \left(\frac{m_n^*-m_p^*
}{m_n^* m_p^*}\right)\nonumber\\
&+&\frac{1}{16}t_3\alpha (2x_3+1)\rho^{\alpha+1}.
\end{eqnarray}
Making the approximation $m_n^* m_p^*\simeq(m_0^*)^2$ one can write the expression 
of $\Delta m_0^*\ \left[=\frac{m_n^*-m_p^*}{m}\right]$ as,
\begin{eqnarray}
\label{delm1}
&&\Delta m_0^*=\left(3C_2^0-L_0-E_F^0+\frac{2}{3}\frac{m}{m_0^*}E_F^0\right.\nonumber\\
&-&\left.\frac{1}{16}t_3\alpha (2x_3+1)\rho^{\alpha+1}\right)\Bigg/\left[E_F^0\left(
\frac{m}{m_0^*}\right)^2\right]
\end{eqnarray}
Using the expression of $K_{sym}^0$ (see Eq. (\ref{ksym2})) to eliminate $t_3$ and 
$x_3$ in Eq. (\ref{delm1}), one obtains
\begin{eqnarray}
\label{}
&&\Delta m_0^*=\Bigg(K_{sym}^0+3(1+\alpha)(3C_2^0-L_0)+(1-3\alpha)E_F^0\nonumber\\
&&+\frac{2}{3}(3\alpha-2)\frac{m}{m_0^*}E_F^0\Bigg)\Bigg/\left[(3\alpha-2)E_F^0\left(
\frac{m}{m_0^*}\right)^2\right].
\end{eqnarray}

%\bibliography{plb_ref}

\end{document}